\documentclass[prc,twocolumn,superscriptaddress,showpacs,floatfix]{revtex4}
\usepackage{graphicx,amsmath,amssymb,bm}

\usepackage{braket}
\usepackage{amsfonts}
\usepackage{subfigure}

\newcommand{\cre}[1]{a^\dagger_{#1}}
\newcommand{\crebar}[1]{\overline{a^\dagger_{#1}}}
\newcommand{\ann}[1]{a_{#1}}
\newcommand{\annbar}[1]{\overline{a_{#1}}}
\newcommand{\anntilde}[1]{\tilde a_{#1}}
\newcommand{\barH}{\overline{H}}

\newcommand{\nuc}[2]{\(^{#1}\)#2}

\newcommand{\CG}[2]{\left(#1\mid#2\right)}
\newcommand{\tProd}[2]{\left[#1\otimes#2\right]}

\begin{document}

\title{Computation of spectroscopic factors with the coupled-cluster method}

\author{\O.~Jensen}
\affiliation{Department of Physics and Technology, University of Bergen,
N-5007 Bergen, Norway}
\author{G.~Hagen}
\affiliation{Physics Division, Oak Ridge National Laboratory,
Oak Ridge, TN 37831, USA}
\author{T.~Papenbrock}
\affiliation{Department of Physics and Astronomy, University of
Tennessee, Knoxville, TN 37996, USA}
\affiliation{Physics Division, Oak Ridge National Laboratory,
Oak Ridge, TN 37831, USA}
\author{D.J.~Dean}
\affiliation{Physics Division, Oak Ridge National Laboratory,
Oak Ridge, TN 37831, USA}
\author{J.S.~Vaagen}
\affiliation{Department of Physics and Technology, University of Bergen,
N-5007 Bergen, Norway}

\begin{abstract}

  We present a calculation of spectroscopic factors within
  coupled-cluster theory. Our derivation of algebraic equations for
  the one-body overlap functions are based on coupled-cluster
  equation-of-motion solutions for the ground and excited states of
  the doubly magic nucleus with mass number $A$ and the odd-mass
  neighbor with mass $A-1$.  As a proof-of-principle calculation, we
  consider $^{16}$O and the odd neighbors $^{15}$O and $^{15}$N, and
  compute the spectroscopic factor for nucleon removal from $^{16}$O.
  We employ a renormalized low-momentum interaction of the
  $V_{\mathrm{low-}k}$ type derived from a chiral interaction at
  next-to-next-to-next-to-leading order. We study the sensitivity of
  our results by variation of the momentum cutoff, and then discuss the
  treatment of the center of mass.

\end{abstract}


\maketitle

\section{Introduction \label{secIntro}}

In the past two decades, {\it ab initio} nuclear structure
calculations have led to the development and test of high-precision
models with predictive
power~\cite{Navratil2000gs,Pieper2001mp,Nogga03}.  Recently, the
application of effective field theory
(EFT)~\cite{vKolck,vKolckRev,Epelbaum,EntemMachleidt-2003} and
renormalization group techniques~\cite{bogner-2003,bogner2009}
resulted in a model-independent approach to the nuclear interaction.
These approaches have significantly deepened our understanding of
nuclear forces and have also provided us with new technical means to
simplify the solution of the nuclear many-body problem. The
interactions from chiral EFT have been probed in light
nuclei~\cite{Nav07,Nogga,NCSMRev,DeanLee} and selected medium-mass
nuclei with different
techniques~\cite{HPD+08,Barbieri2009prl,Fujii2009}. The focus of {\it
  ab initio} calculations is not only on observables such as binding
energies, radii, and low-lying excitation spectra, but also on
transition rates and more detailed spectroscopic information. Very
recently, {\it ab initio} theory began to bridge the gap from nuclear
structure to reactions~\cite{hagen-2007plb,Nollet,quaglioni08}. The
inclusion of continuum effects, for instance, is necessary for the
description of weakly bound and unbound nuclei.  Direct reactions such
as stripping and pickup of a single nucleon are rather well understood
within phenomenological approaches (see, e.g. Ref.~\cite{hansen}), but
constitute a current frontier for {\it ab initio} theory.

The interpretation of direct reactions within a given model or
Hamiltonian is based on spectroscopic
factors~\cite{Macfarlane,bang1985one}.  The spectroscopic factor
depends on wave function overlaps (see
Eq.~(\ref{equOverlapInvariantDef}) below for a definition) and
provides useful information that relates nuclear structure within a
given model (i.e. within a given Hamiltonian) to stripping and
transfer reactions~\cite{Macfarlane}.  The spectroscopic factor is not
an observable as it depends on the employed Hamiltonian or model.  In
nuclear physics, the high-momentum parts of the interaction are
unconstrained and modeled in different ways. Thus, the short-ranged
part of the wave function is model-dependent, and so is an overlap
between wave functions.  Therefore, the spectroscopic factor is merely
a theoretical quantity and cannot be
measured~\cite{Furnstahl-2002,furnstahl-2010}.  However, the
spectroscopic factor ``provides a useful basis for the comparison of
experiment and current nuclear models''~\cite{Macfarlane}. Its purpose
thus lies in understanding a direct reaction within a certain model or
Hamiltonian, and this interpretation might be useful and
interesting~\cite{Barbieri2009spe}.

In this paper, we develop the technical tools to compute spectroscopic
factors within the coupled-cluster
method~\cite{Coester,CK,Cizek,Cizek2,KLZ} (see
Ref.~\cite{bartlett2007} for a recent review of this method), and
perform a proof-of-principle calculation for $^{16}$O. The computation
of the spectroscopic factor within coupled-cluster theory is not
trivial (i) since the method does not readily yield a many-body wave
function, and (ii) due to details related to the translation
invariance of the coupled-cluster wave function.  This paper is
structured as follows. Section \ref{secTheory} is dedicated to a
summary of the employed coupled-cluster method.  The theoretical
computation of spectroscopic factors within coupled-cluster theory is
presented in Section~\ref{secSF}. We present our results and a
discussion of the center-of-mass treatment in Section~\ref{secRes}.
Section \ref{secCnO} contains our conclusions and an outlook.

\section{Equation-of-motion and coupled-cluster theory for nuclei}
\label{secTheory}

In this section we introduce the Hamiltonian and coupled-cluster
theory~\cite{Coester,CK,Cizek,Cizek2,KLZ} for closed-shell and
open-shell nuclei. Although our implementation of coupled-cluster theory
has been presented elsewhere \cite{bartlett2007, dean2004,
  hagen2007-bench, HPD+08, hagen2009}, we give a brief overview of the
method since some details are needed for the calculation
of spectroscopic factors.

We consider the intrinsic nuclear $A$-body Hamiltonian
\begin{eqnarray}
\nonumber
\hat{H} &=& \hat{T}-\hat{T}_{\rm cm} +\hat{V}\\
&=& \sum_{1\le i<j\le A} {(\vec{p}_i-\vec{p}_j)^2\over 2mA} +\hat{V} \ .
\label{equHamiltonOperatorA}
\end{eqnarray}
Here $T$ is the kinetic energy, $T_{\rm cm}$ is the kinetic energy of
the center-of-mass coordinate, and $V$ is the two-body nucleon-nucleon
interaction.  In this paper we use low-momentum interactions $V_{{\rm
    low-}k}$ \cite{Bogner2003, bogner2009} with sharp cutoffs $\lambda
= 1.6, 1.8, 2.0, 2.2~\mathrm{fm}^{-1}$, respectively. For simplicity,
we neglect any contributions of three-nucleon forces as we focus on a
proof-of-principle calculation.

In coupled-cluster theory, one writes the ground-state many-body wave
function as
\begin{equation}
	\ket{\psi_0} = e^T\ket{\phi_0} \ .
	\label{equCCAnsatz}
\end{equation}
Here, $\ket{\phi_0}$ is a product state.
The cluster operator $T$ introduces correlations as
a linear combination of particle-hole excitations
\begin{equation}
	T = T_1 + T_2 + \ldots + T_A \ .
	\label{equClusterOperatorDef}
\end{equation}
Here, the $n$-particle-$n$-hole excitation operator is
\begin{equation}
	T_n = \left(\frac{1}{n!}\right)^2\prod_{\nu=1}^n\sum_{a_\nu,i_\nu}
	t_{i_1\cdots i_n}^{a_1\cdots a_n} \cre{a_1}\cdots\cre{a_n}\ann{i_n}\cdots\ann{i_1} \ .
	\label{equExcitationOperatorDef}
\end{equation}
We employ the standard convention that indices $ijk\ldots$ refer to
orbits below Fermi level (holes) and $abc\ldots$ above Fermi level
(particles).  Approximations in coupled-cluster theory are introduced
by truncating the cluster operator $T$ at a certain particle-hole
excitation level.  In this work we truncate $T$ at the
two-particle-two-hole excitation level, i.e.  $T\approx T_1+T_2$,
which gives the coupled-cluster method with singles and doubles
excitations (CCSD). This is the most commonly used approximation, as it
provides a good compromise between computational cost on the one hand
and accuracy on the other.

Within the CCSD approximation, the computational cost is given by
$n_o^2 n_u^4$, where $n_o$ and $n_u$ denote the number of occupied and
unoccupied orbitals, respectively.  The correlated ground state
solution is given by the amplitudes $t_i^a$ and $t_{ij}^{ab}$ that
solve the nonlinear equations
\begin{eqnarray}
  \label{ccsd}
  \braket{\phi_i^a|\barH|\phi_0} &=& 0 \ ,  \\
  \braket{\phi_{ij}^{ab}|\barH|\phi_0} &=& 0 \ .
\end{eqnarray}
Here, the bra states are particle-hole excitations of the reference
Slater determinant, and $\barH$ denotes the similarity-transformed
Hamiltonian,
\begin{equation}
  \barH = e^{-T}He^T = \left(He^T\right)_c \ .
  \label{equHbarDefinition}
\end{equation}
The subscript $c$ indicates that only fully connected diagrams give
non-zero contributions.  Once $T$ is determined from the solution of
the coupled-cluster equations~(\ref{ccsd}), the correlated ground
state (g.s.)  energy is given by
\begin{equation}
	E_0 = \braket{\phi_0|\barH|\phi_0} \ .
	\label{equCCenergy}
\end{equation}
The CCSD approach is known to work particularly well for the ground
state of nuclei with closed (sub-) shells, as a Slater determinant provides a
reasonable first approximation.  In this work we use the Equation-of-Motion
(EOM) \cite{kowalski2004, gour2006, bartlett2007, Geertsen1989, Rowe1968}
method to solve for the ground and excited states of the closed-shell nucleus
$A$ and its odd neighbors with mass number $A - 1$.

In EOM, the ground and excited states of a nucleus with mass number
$B$ are obtained by acting with an excitation operator $\Omega_\mu$ on
the ground state wave function of a nucleus with mass number $A$, i.e.
$\psi^B_\mu = \Omega_\mu \psi_0^A$. Here $\mu$ denotes quantum numbers
such as spin, parity, and isospin projection.  Within the
EOM approach, the ground state
wave function $\psi_0^A$ denotes the coupled-cluster wave
function $e^T\phi_0$.  In this work we choose either $B=A$, in which case we solve
the excited states of closed-shell nucleus $A$, or $B=A-1$, in which
case we solve the ground and excited states of the $A-1$ neighboring
nucleus.  To solve for the excited states of the closed-shell nucleus
$A$, we define $\Omega_\mu$ by the excitation operators,
\begin{eqnarray}
	R^A &=& r_0 +  \sum_{ia} r^a_i \cre{a}\ann{i}
			+ \frac14 \sum_{ijab} r^{ab}_{ij}\cre{a}\cre{b}\ann{j}\ann{i}\ ,\\
	L^A &=& 1 + \sum_{ia} l_a^i \cre{i}\ann{a}
			+ \frac14 \sum_{ijab} l^{ij}_{ab}\cre{i}\cre{j}\ann{b}\ann{a}\ .
\end{eqnarray}
Here, we suppressed the index $\mu$, but it is understood that the
operators $R^A$ and $L^A$ excite and de-excite states with quantum
numbers $\mu$, respectively. For the ground and excited states of
the nucleus with mass number ${A-1}$, we define $\Omega_\mu$ by the particle removal
operators
\begin{eqnarray}
	R^{A-1} &=& \sum_i r_i \ann{i} + \frac12 \sum_{ija} r^{a}_{ij}\cre{a}\ann{j}\ann{i}\ ,  \\
	L^{A-1} &=& \sum_i l^i \cre{i} + \frac12 \sum_{ija} l_{a}^{ij}\cre{i}\cre{j}\ann{a}\ .
\end{eqnarray}
Again, we supressed the index $\mu$ labeling the quantum numbers. The
operators $R_\mu =R^{A}$ ($R_\mu=R^{A-1}$) commute with the cluster operator $T$, and
the unknowns $r_i^a, r_{ij}^{ab}$ ($r_i, r_{ij}^a$) solve the EOM equation
\begin{equation}
\left[ \overline{H}, R_\mu \right] \vert \phi_0 \rangle = \omega_\mu
R_\mu \vert \phi_0 \rangle \ ,
\label{eq:paeqn2}
\end{equation}
which defines an eigenvalue problem for the excitation operator
$R_\mu$ with eigenvalue $\omega_\mu = E_\mu-E_0$.  It is clear from
the definitions (\ref{equExcitationOperatorDef}) and
(\ref{equHbarDefinition}) that $\barH$ is non-Hermitian; dual
space solutions need to be calculated explicitly.  We obtain the
de-excitation operators $L_\mu =L^{A},L^{A-1}$ by solving the
left eigenvalue problem
\begin{equation}
  \bra{\phi_0} L_\mu \barH = \bra{\phi_0} L_\mu \omega_\mu \ .
  \label{equEOMCC}
\end{equation}
The right and left eigenvectors form a bi-orthogonal set and are
normalized in the following way
\begin{equation} \label{equNormalizationCondition}
  \bra{\phi_0} L_\mu R_{\mu'} \vert \phi_0\rangle = \delta_{\mu\mu'} \ .
\end{equation}

The EOM solution for the ground state of system $A$ is identical to the CC
solution, so that $R^A_{0} = r_0 = 1$.  Ref~\cite{bartlett2007} provides a
detailed description of EOM.

\section{Overlap functions and spectroscopic factors
  from coupled-cluster theory}\label{secSF}

The one-particle overlap function between two wave functions
$\Psi_{A-1}$ and $\Psi_A$ of nuclei with mass number $A-1$ and $A$,
respectively, is defined as \cite{bang1985one}
\begin{equation}\label{equOverlapInvariantDef}
  O^A_{A-1}(\vec x) \equiv
  \sqrt A \int \mathrm d^{3(A-2)}\xi \Psi_{A-1}^*(\vec \xi) \Psi_A(\vec x,\vec \xi) \ .
\end{equation}
Here $\xi$ represent the $3(A-2)$ translationally invariant position
coordinates and the $A-1$ spin coordinates of $A-1$ particles present
in both $\Psi_{A-1}$ and $\Psi_A$, while $\vec x$ labels the position
and spin of the additional particle in the nucleus with mass number
$A$ with respect to the center of mass of the nucleus with mass number
$A-1$.  The isospin coordinate has been suppressed.

In our coupled-cluster approach, however, we do not employ coordinates
with respect to the center of mass, as this would limit us to light
systems~\cite{Bishop1990}. Thus, the overlap can be associated with a specific
nucleon represented by a second quantization operator,
\begin{equation} \label{equOverlapDef}
	O^A_{A-1}(\vec x) \equiv \braket{{A-1}|\ann{}(\vec x)|A} \ .
\end{equation}
Here, $|A\rangle$ and $|A-1\rangle$ denote eigenstates in the nucleus
with mass $A$ and $A-1$, respectively. Typically, $|A\rangle$ is the
ground state, and $|A-1\rangle$ is the ground state or an excited
state. Our formalism will be kept general and is not limited to these
cases.  The radial overlap function $O_{A-1}^A(lj;r)$ is derived by
expanding $\vec x$ in terms of partial waves,
\begin{equation}
	\ann{}(\vec x) = (-)^{j-m}\sum_{ljm}\anntilde{lj{-m}}(r)Y_{ljm}(\hat x) \ .
\end{equation}
Here, $Y_{ljm}(\hat x)$ is the spin-orbital spherical harmonic
\[
Y_{ljm}(\hat x) = \tProd{Y_{l}(\hat{r})}{\chi_{1/2}(\sigma)}_{jm} \ ,
\]
$Y_{l}(\hat r)$ is the spherical harmonic of rank $l$ and
$\chi_{1/2}(\sigma)$ is a fermionic spin function.  The orbital
angular momentum quantum number is denoted by $l$, while $j$ and $m$
denote the rank and projection, respectively, of $Y_{ljm}(\hat x)$ as
a spherical tensor.  The hat denotes unit vectors, i.e. $\hat{x}\equiv
\vec{x}/|\vec{x}|$.  We have also introduced the spherical
annihilation operator $\anntilde{ljm}(r) = (-)^{j+m}\ann{lj{-m}}(r)$.
The radial overlap is now given by the reduced matrix element and the
overlap becomes \footnote{Many authors use an alternative definition
  derived from $\braket{A|\cre{}(\vec x)|{A-1}}$}
\begin{multline}
  O_{A-1}^A(\vec x) = \sum_{j}(-)^{j-m} \CG{J_AM_Aj-m}{J_{A-1}M_{A-1}} \\
  \times O_{A-1}^A(lj;r) Y_{ljm}(\hat x) \ .
  \label{equSphericalOverlapDef}
\end{multline}
Here, $\CG{\cdot}{\cdot}$ denotes a Clebsch-Gordan coefficient.  The
overlap is now expressed by a radial function associated with each
tensorial component $Y_{ljm}(\hat x)$
\begin{eqnarray}			\label{equOverlapDefSQ}
	O_{A-1}^{A}(lj;r) &\equiv& \braket{{A-1}||\anntilde{lj}(r)||A}\\
	&=& (-)^{j-m}\frac{\braket{{A-1}M_{A-1}|\ann{ljm}(r)|AM_A}}
{\CG{J_AM_Aj{-m}}{J_{A-1}M_{A-1}}} \ . \nonumber
\end{eqnarray}
This equation also defines the reduced matrix elements we employ.

The norm of the radial overlap function is the spectroscopic factor
\begin{equation}
	S_{A-1}^A(lj) =  \int \mathrm dr r^2 \left|O_{A-1}^A(lj;r)\right|^2 \ .
	\label{equSFdef}
\end{equation}
The overlap functions can be expressed in an energy basis by inserting the
expansions
\begin{eqnarray}
	\cre{ljm}(r) &=& \sum_n \cre{nljm} \phi_{nlj}(r) \ ,
				\label{equCreationOperatorNodalExpansion} \\
	\ann{ljm}(r) &=& \sum_n \ann{nljm} \phi^*_{nlj}(r) \ ,
				\label{equAnnihilationOperatorNodalExpansion}
\end{eqnarray}
where $n$ is the nodal quantum number and $\phi_{nlj}(r)$ is the
radial single-particle wave function associated with the orbits
$nljm$.  While $\cre{ljm}(r)$ represents the creation of a particle at
radial distance $r$, $\cre{nljm}$ represents the action of populating
a single-particle orbit.

Assuming orthogonality of the single-particle wave functions, the
spectroscopic factor is written as,
\begin{eqnarray}
  \nonumber
  S_{A-1}^A(lj) & = &
  \sum_{n} \vert \braket{{A-1}||\anntilde{nlj}||A} \vert^2  \\
  & = & \sum_n\frac{ \vert \braket{A-1|\ann{nljm}|A} \vert^2 }
  {\CG{J_AM_Aj{-m}}{J_{A-1}M_{A-1}}^2} \ .
  \label{equSFdefinition}
\end{eqnarray}
Here we used the Wigner-Eckhart theorem for the reduced matrix
elements.  Eq.~(\ref{equSFdefinition}) is our starting point since we
work in an uncoupled ($m-$scheme) basis.  Using the EOM-CCSD solutions
for the right and left eigenvalue problems for the $A$ and the $A-1$
systems, and employing the ground state solutions for system $A$,
Eq.~(\ref{equSFdefinition}) takes the form,
\begin{multline}
  S_{A-1}^A(lj) =\\
	\sum_n \frac{
		\braket{\phi_0 | L_0^A \crebar{nljm} R_\mu^{A-1} |\phi_0}
		\braket{\phi_0 | L_\mu^{A-1} \annbar{nljm} R_0^A |\phi_0}
		}
  {\CG{J_AM_Aj{-m}}{J_{A-1}M_{A-1}}^2} \ .
  \label{equSFdefinitionCC}
\end{multline}
This gives the equation for the spectroscopic factors as defined
within coupled-cluster theory. We note that this equation is
unambiguously and uniquely defined in terms of the left and right
eigenstates of the nuclei with mass numbers $A$ and $A-1$.  This is
clear since the spectroscopic factor is given by the absolute value
squared of the one-body overlap matrix element, so any ambiguity
related to the normalization condition
(\ref{equNormalizationCondition}) is removed.

In Eq.~(\ref{equSFdefinitionCC}) we have introduced the
similarity-transformed creation and annihilation operators,
\begin{eqnarray}
	\crebar{p} &=& e^{-T}\cre{p} e^T \ ,\\
	\annbar{p} &=& e^{-T}\ann{p} e^T \ .
\end{eqnarray}
Using the Baker-Campbell-Hausdorff commutator expansion, we can derive
algebraic expressions for $\crebar{p}$ and $\annbar{p}$ in terms of
the ``bare'' creation and annihilation operators $\cre{p}$ and
$\ann{p}$ and the particle-hole excitations amplitudes $t_i^a$ and
$t_{ij}^{ab}$,
\begin{eqnarray}
	\label{equHaussdorfCrebar}
	\crebar{p} &=& \cre{p} - \sum_{b} t_p^b \cre{b}
			- \frac12 \sum_{jbc} t_{pj}^{bc} \cre{b}\cre{c}\ann{j} \ , \\
	\label{equHaussdorfAnnbar}
	\annbar{p} &=& \ann{p} + \sum_{i} t_i^p \ann{i}
	+ \frac12 \sum_{ijc} t_{ij}^{pc}\cre{c}\ann{j}\ann{i} \ .
\end{eqnarray}
These equations can also be given in a diagrammatic form, which
provides a convenient bookkeeping system for the available Wick
contractions in the expressions.

The coupled-cluster diagrams are similar to Goldstone diagrams. An
algebraic Wick contraction corresponds to the diagrammatic connection
of two directed lines, but the interpretation rules are slightly
different.  We refer the reader to Refs.~\cite{crawford2000,
  bartlett2007} for a complete introduction to the diagrammatic
approach.  Here we will only present the few concepts necessary
to introduce the novel extensions of the formalism used in the context
of spectroscopic factors.

\begin{table}[tb]
  \begin{tabular}{r|l}
    $T_1$ & \includegraphics[scale=0.1]{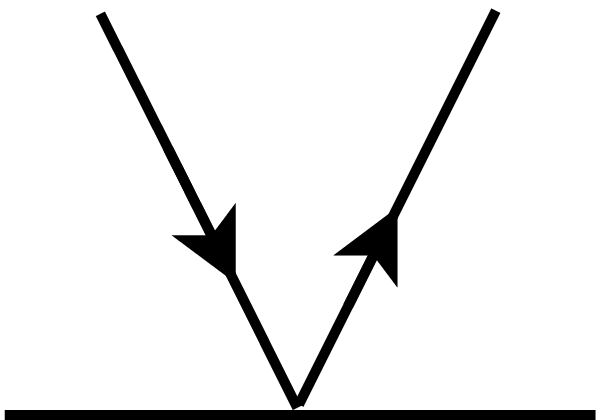} \\
    $T_2$ & \includegraphics[scale=0.1]{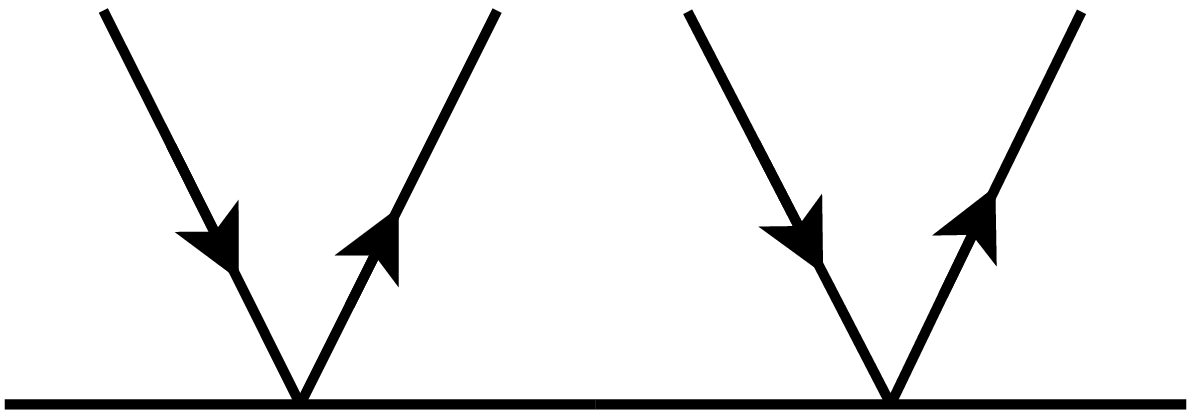} \\
    $L^A$ & \includegraphics[scale=0.1]{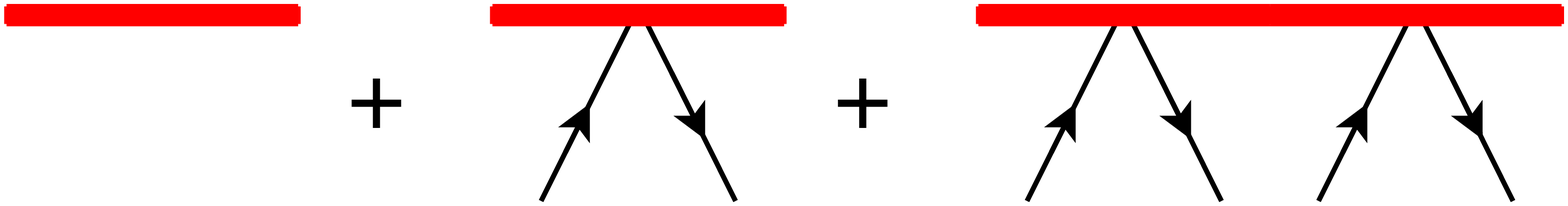} \\
    $L^{A-1}$ & \includegraphics[scale=0.1]{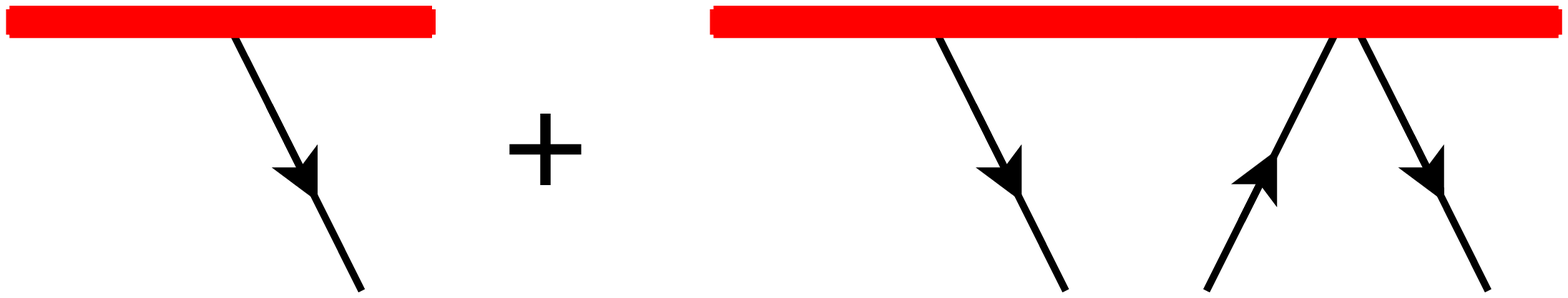} \\
    $R^A$ & \includegraphics[scale=0.1]{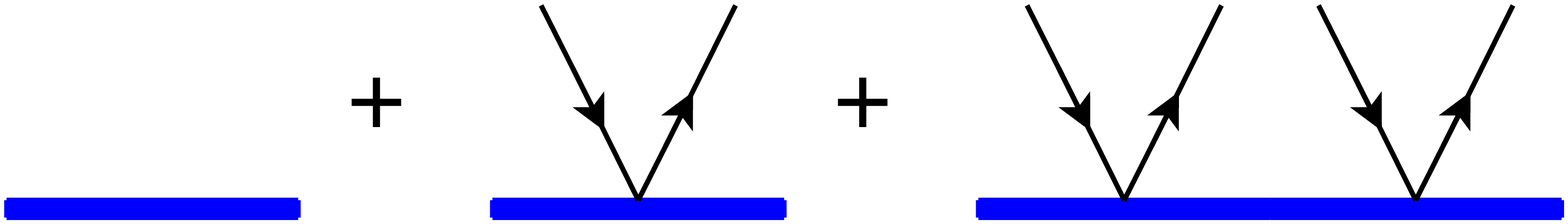} \\
    $R^{A-1}$ & \includegraphics[scale=0.1]{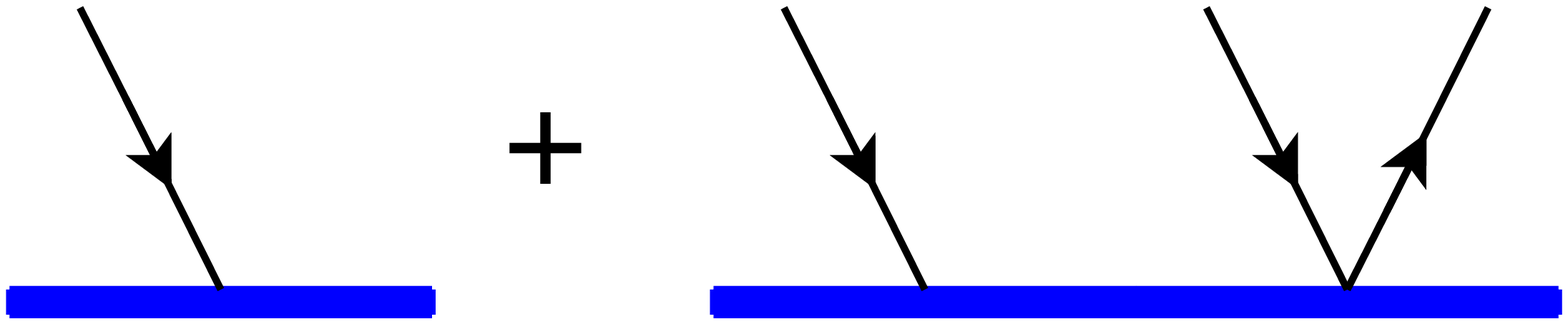}
  \end{tabular}
  \caption{
    The left panel lists algebraic symbols for the particle-hole excitation
    operators and particle removal operators.  Corresponding diagrammatic
    representations are displayed on the right.  Arrows pointing up (down)
    represent particle (hole) orbits with implicit summation indices
    $a,b,c,\ldots$ ($i.j,k,\ldots$).  \label{figDiagramsTLR}
  }
\end{table}

Diagrammatic representations of the excitation and
particle-removal operators $T$, $L_\mu$, and $R_\mu$ are displayed in
Table \ref{figDiagramsTLR}.  Lines with arrows pointing up (down)
represent particle (hole) orbits.  These lines have implicit indices
$a,b,c,\ldots$ ($i.j,k,\ldots$) that are summed over.  We suppress
both the summation symbol and the dummy indices for a cleaner
notation.

We have to deal with diagrams that represent operators with an index
that is not being summed over. Such a creation (annihilation) operator
is represented by a directed line pointing out from (in to) a small
circular vertex.  The corresponding diagrams are displayed in the
upper half of Table \ref{figDiagramsOperators}.

Eqs.~(\ref{equHaussdorfCrebar}) and (\ref{equHaussdorfAnnbar}) can be
reproduced diagrammatically as displayed in the lower half of
Table~\ref{figDiagramsOperators}.  The possible Wick contractions
between the creation and annihilation operators $a_p^\dagger$ and
$a_p$, and the cluster operators $T_1$ and $T_2$ depend on whether the
index $p$ denotes an orbital above or below the Fermi surface.  The
small circular vertices distinguish the index fixed by the operator
and is not summed over. In practice, the circle prevents an
accidental connection of the operator line, which would have
introduced an erroneous Wick contraction when the spectroscopic factor
diagrams are written down.

The overall sign of a diagram is determined according to standard
rules~\cite{crawford2000}. The negative sign of the second and third
term in Eq.~(\ref{equHaussdorfCrebar}) is reflected in the
$\crebar{i}$ diagrams by the internal hole-lines that connect the
small circle with the $T$ operators.  To determine the overall sign
correctly for the spectroscopic factor diagrams, a sequence of
directed lines ending or starting in a small circular vertex must be
counted as a loop.

The diagrams shown in Tables~\ref{figDiagramsTLR} and
\ref{figDiagramsOperators} are the basic building blocks for the
computation of the spectroscopic factor.  We compute the matrix
elements of the overlap function as products of the components
$R_\mu$, $L_\mu$ and either $\crebar{p}$ or $\annbar{p}$.  The only
non-vanishing contributions to the spectroscopic factor come from the
diagrams in which all directed lines can be connected.  These diagrams
and the corresponding algebraic interpretation are shown in
Table~\ref{figDiagramsOverlaps}.  We assume an implicit summation over
repeated indices.

The computational cost of the spectroscopic factor diagrams has the very gentle
scaling, $n_o^2n_u^2$, so the cost is completely dominated by the CC and EOM
calculations.  In the case that $|A\rangle$ is the ground state of the closed-shell
nucleus, we have $r_i^a=0=r_{ij}^{ab}$, and several diagrams vanish.

\begin{table}[tb]
  \begin{tabular}{r|l}
    $\cre{i}$, $\cre{a}$ & \includegraphics[scale=0.1]{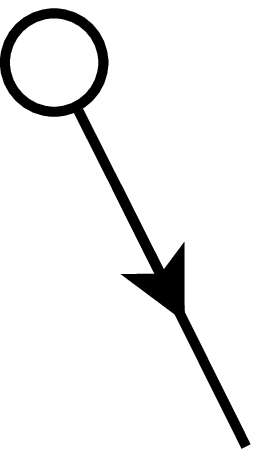}\, ,
    \includegraphics[scale=0.1]{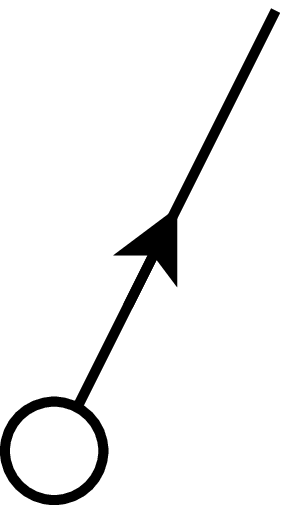} \\
    $\ann{i}$, $\ann{a}$ & \includegraphics[scale=0.1]{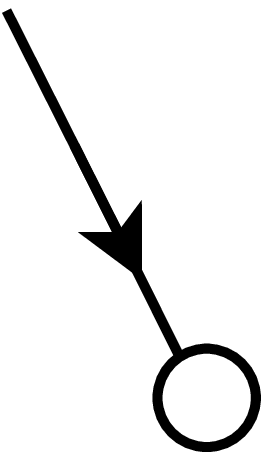}\, ,
    \includegraphics[scale=0.1]{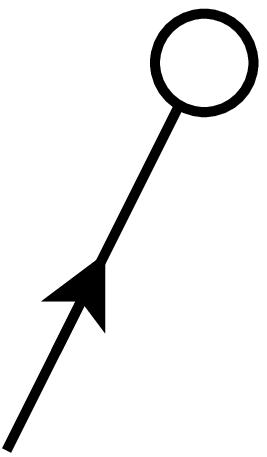} \\
    \hline
    $\crebar{i}$ & \includegraphics[scale=0.1]{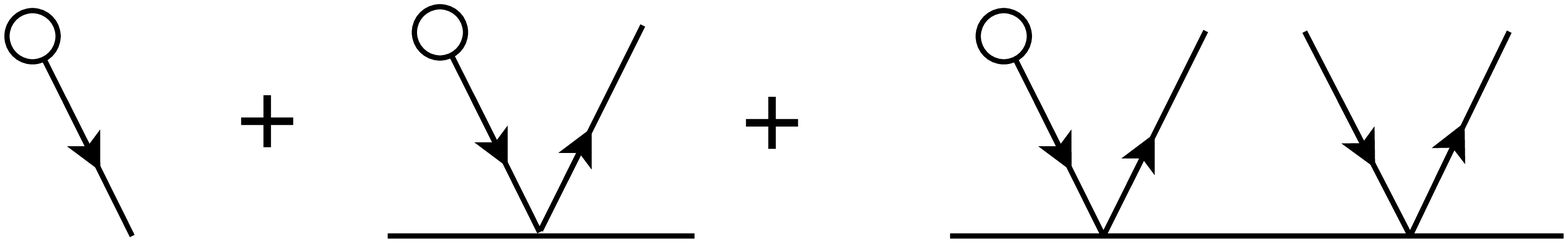} \\
    $\crebar{a}$ & \includegraphics[scale=0.1]{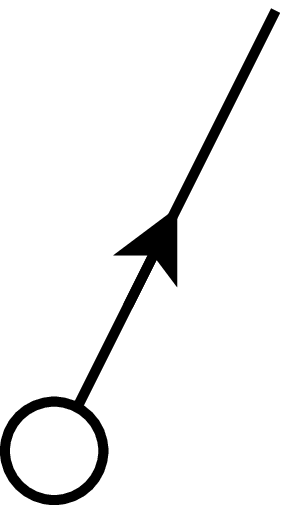} \\
    $\annbar{i}$ & \includegraphics[scale=0.1]{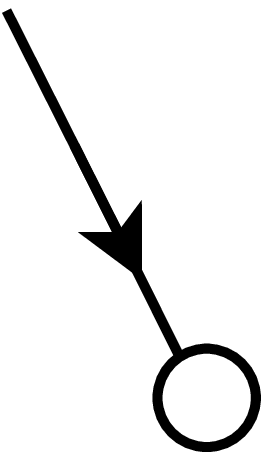} \\
    $\annbar{a}$ & \includegraphics[scale=0.1]{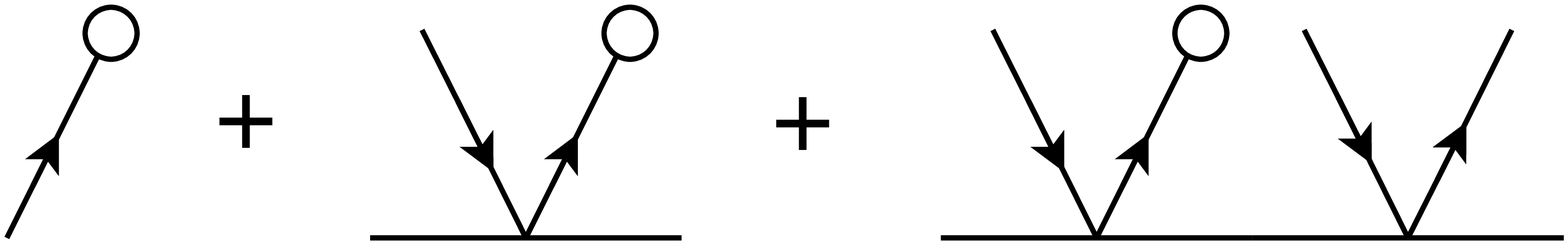} \\
  \end{tabular}
  \caption{
    Diagram representation of the ``bare'' and the similarity-transformed second
    quantization operators.  The black horizontal bars represent the cluster
    operators $T_1$ and $T_2$, as displayed in Table \ref{figDiagramsTLR}.
  }
  \label{figDiagramsOperators}
\end{table}

\begin{table}[tb]
  \begin{tabular}{r|c|c}
    $\braket{{A-1}|\ann{i}|A}$	&\includegraphics[scale=0.1]{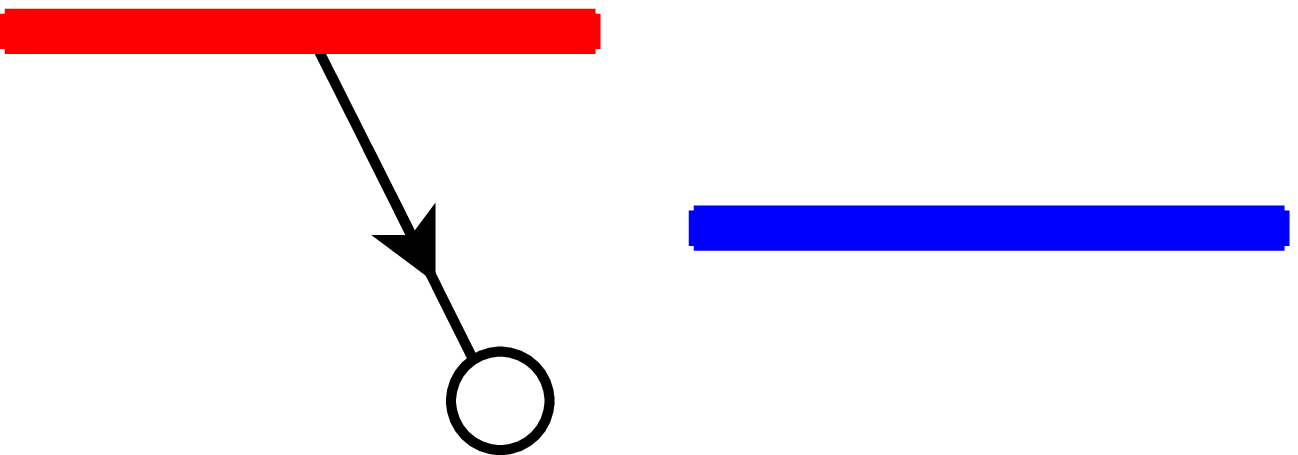} & $l^ir_0$  \\
    & \includegraphics[scale=0.1]{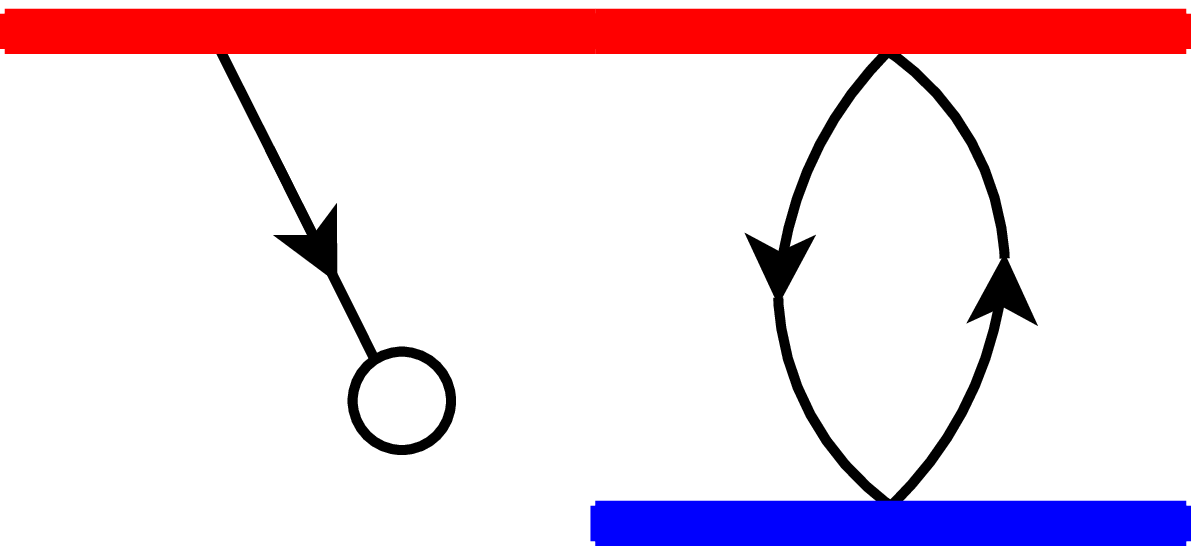} & $l^{ij}_ar_j^a$ \\
    \hline
    $\braket{{A-1}|\ann{a}|A}$	&\includegraphics[scale=0.1]{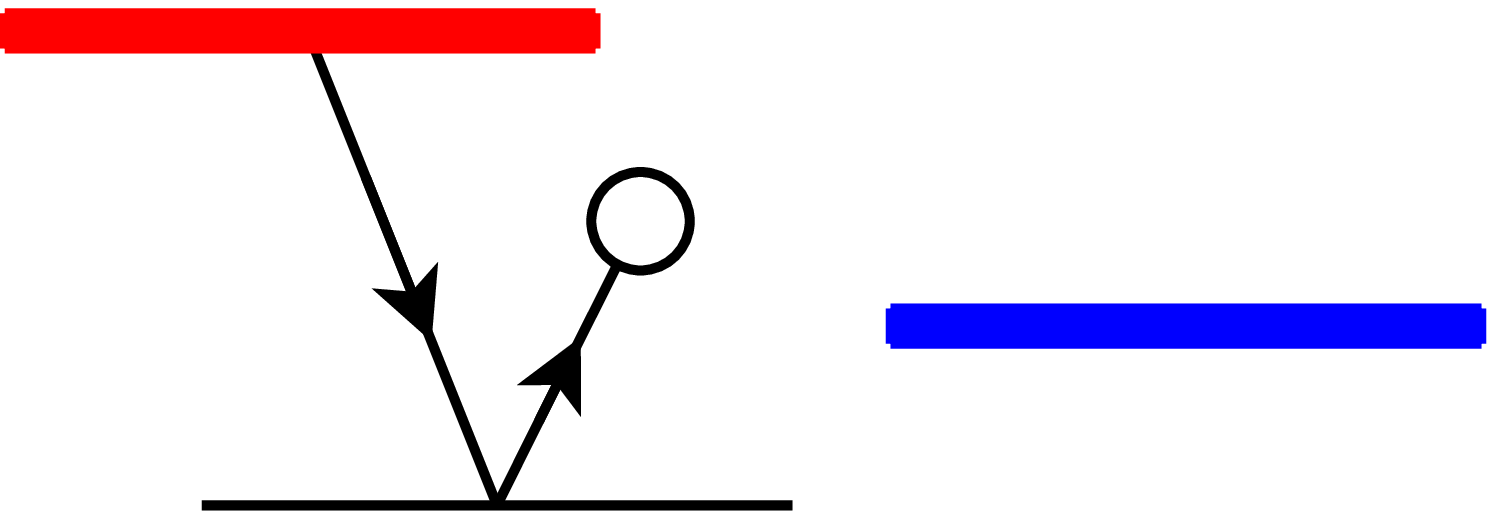} & $l^it_i^ar_0$  \\
    & \includegraphics[scale=0.1]{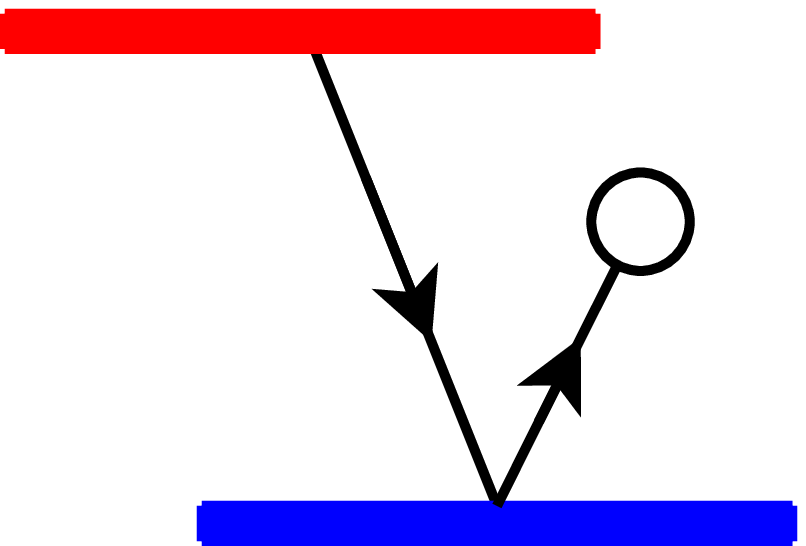} & $l^ir_i^a$  \\
    & \includegraphics[scale=0.1]{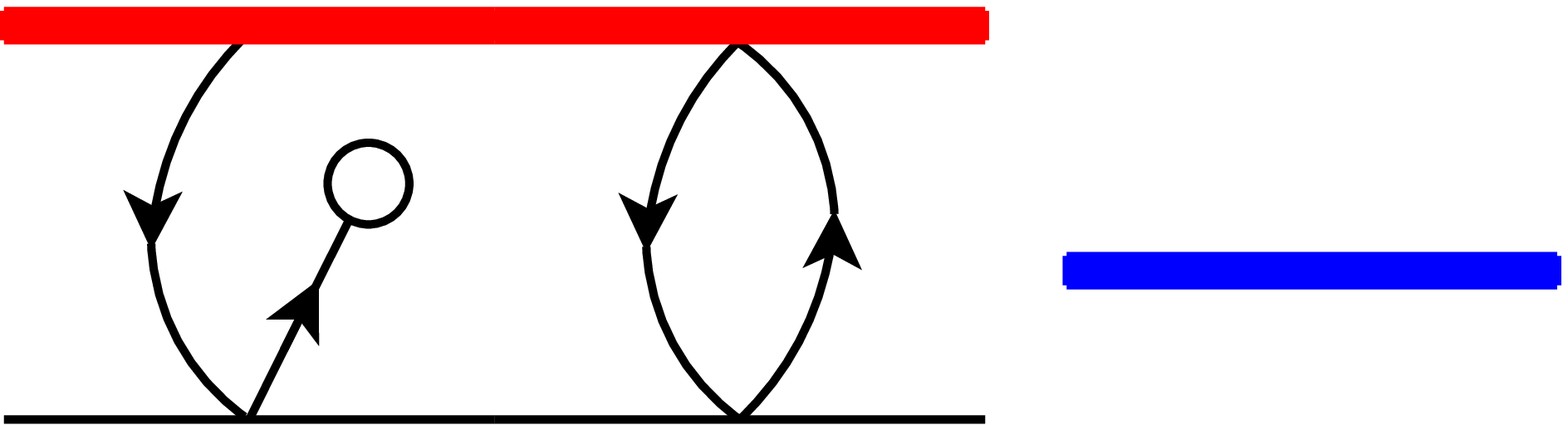} & $\frac12 l^{ij}_bt_{ij}^{ab}r_0$  \\
    & \includegraphics[scale=0.1]{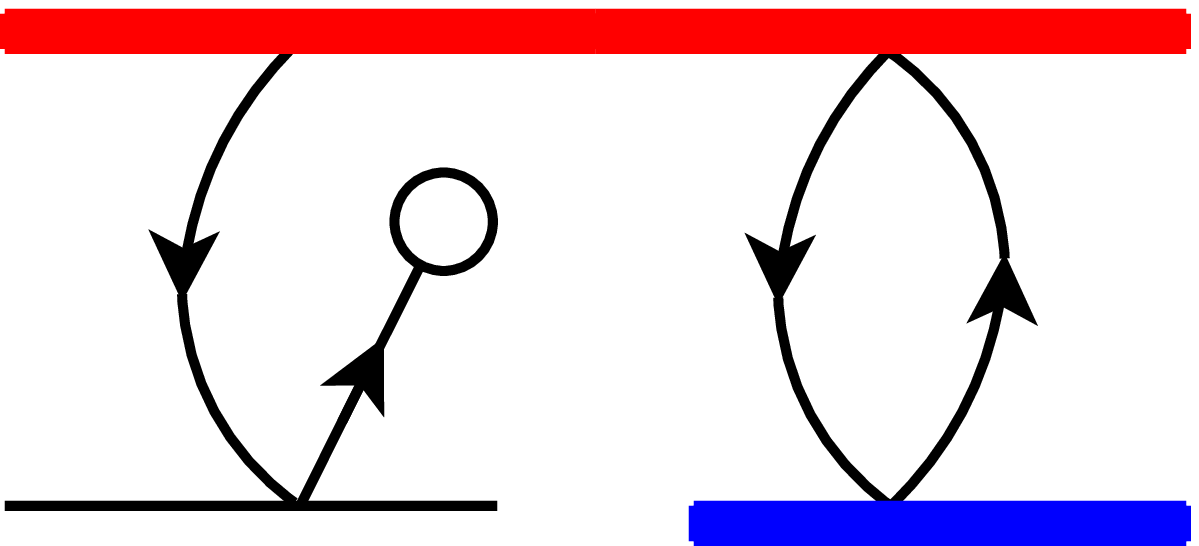} & $l^{ij}_bt_i^ar_j^b$  \\
    & \includegraphics[scale=0.1]{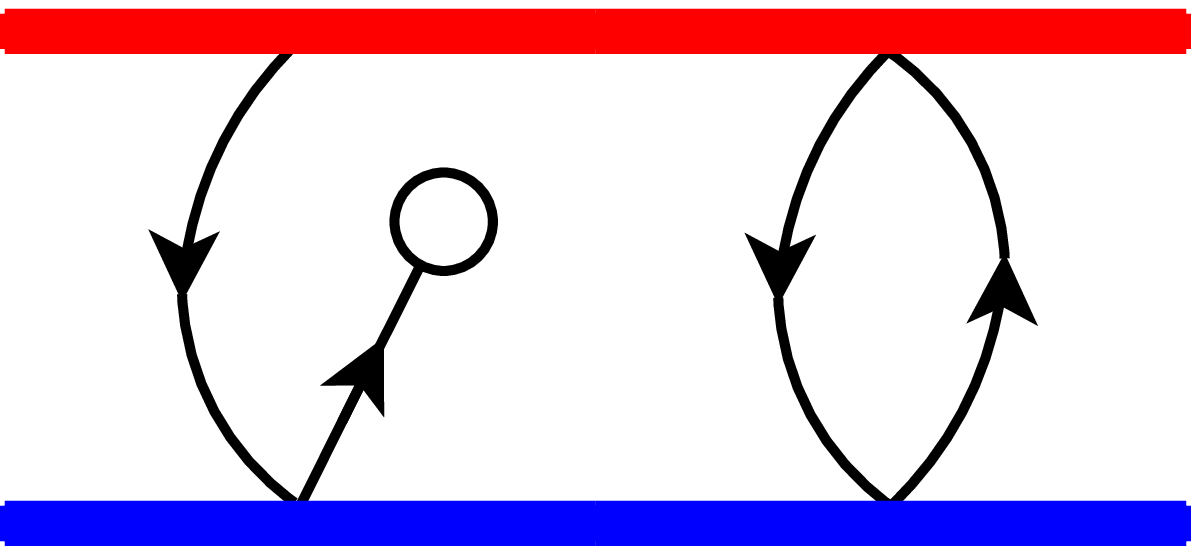} & $\frac12 l^{ij}_br_{ij}^{ab}$  \\
    \hline
    $\braket{A|\cre{a}|{A-1}}$	&\includegraphics[scale=0.1]{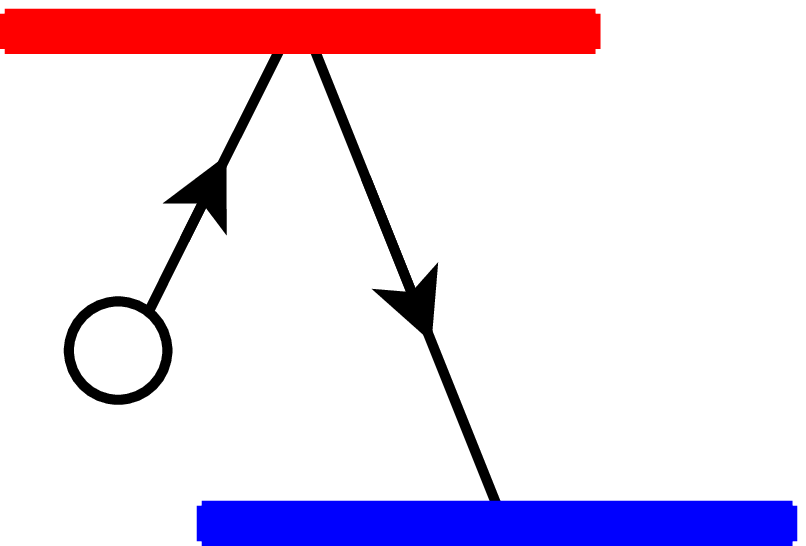} & $l^i_ar_i$  \\
    &\includegraphics[scale=0.1]{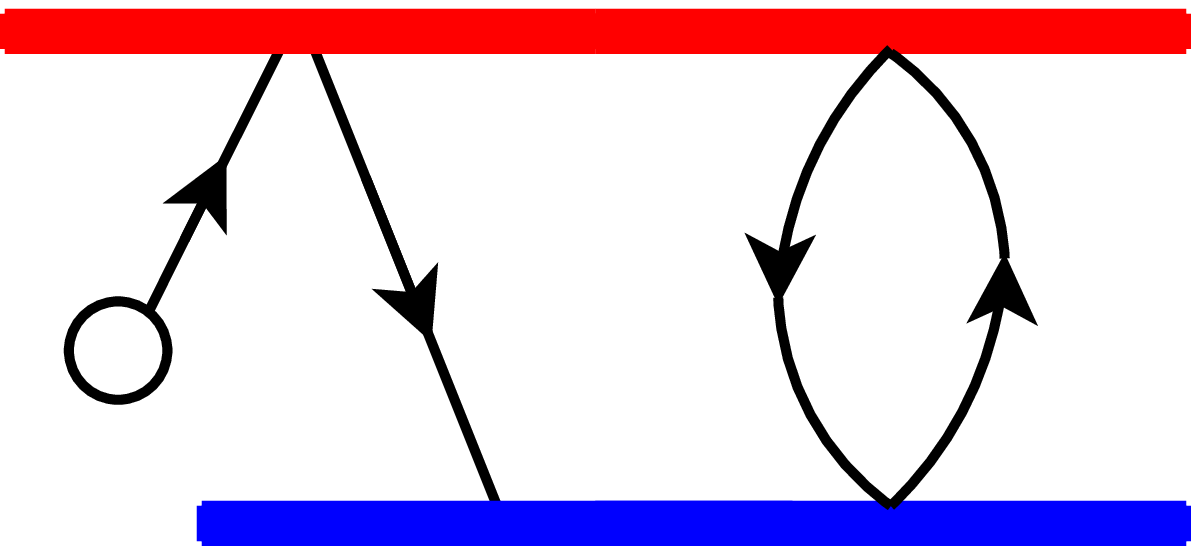} & $\frac12 l^{ij}_{ab}r_{ij}^b$  \\
    \hline
    $\braket{A|\cre{i}|{A-1}}$	&\includegraphics[scale=0.1]{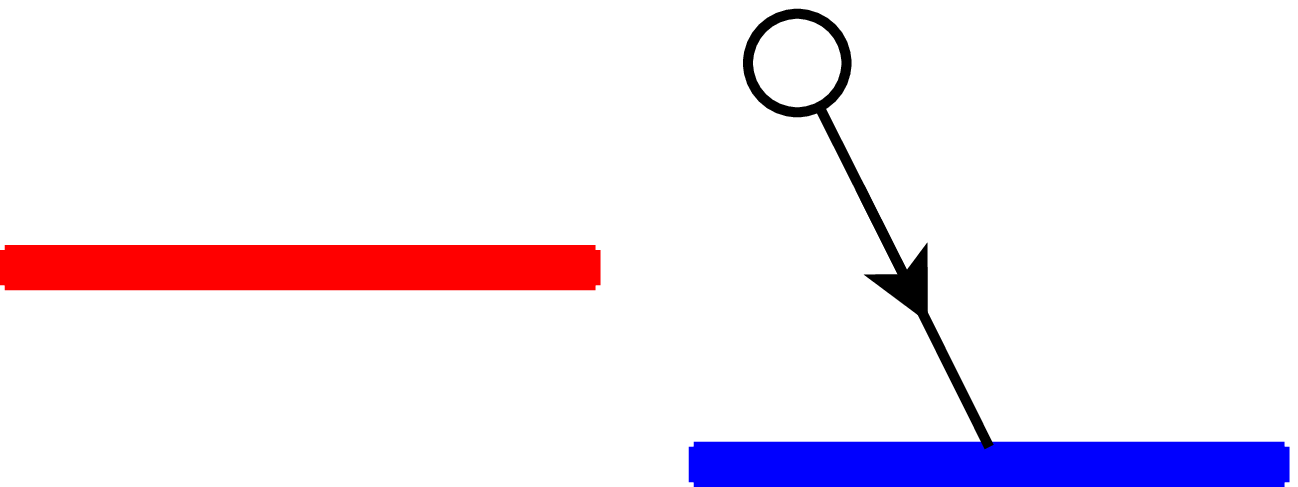} & $l^0r_i$  \\
    &\includegraphics[scale=0.1]{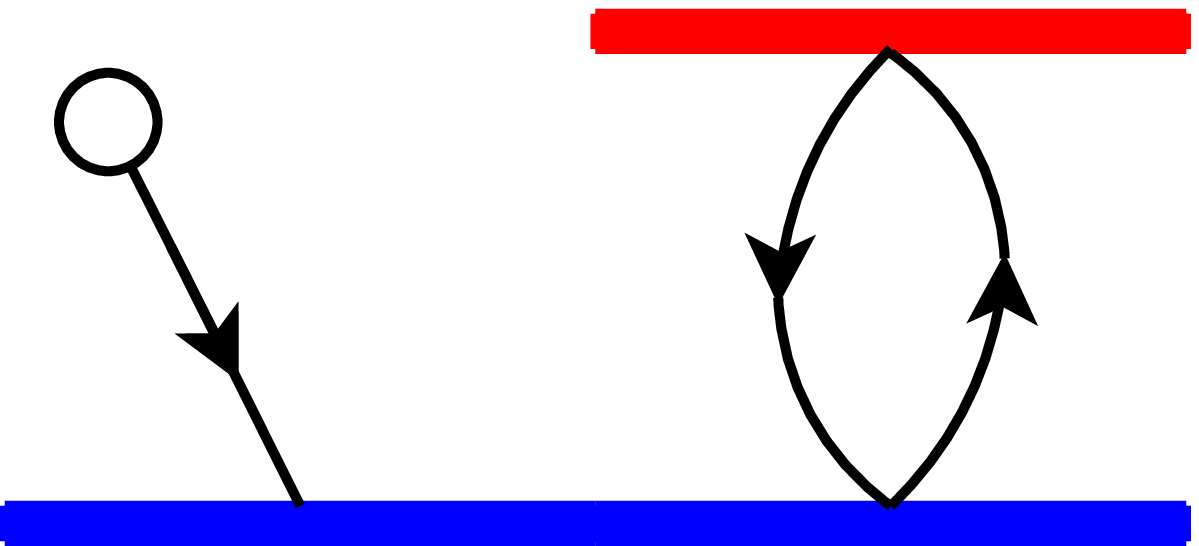} & $l_a^jr_{ij}^a$  \\
    &\includegraphics[scale=0.1]{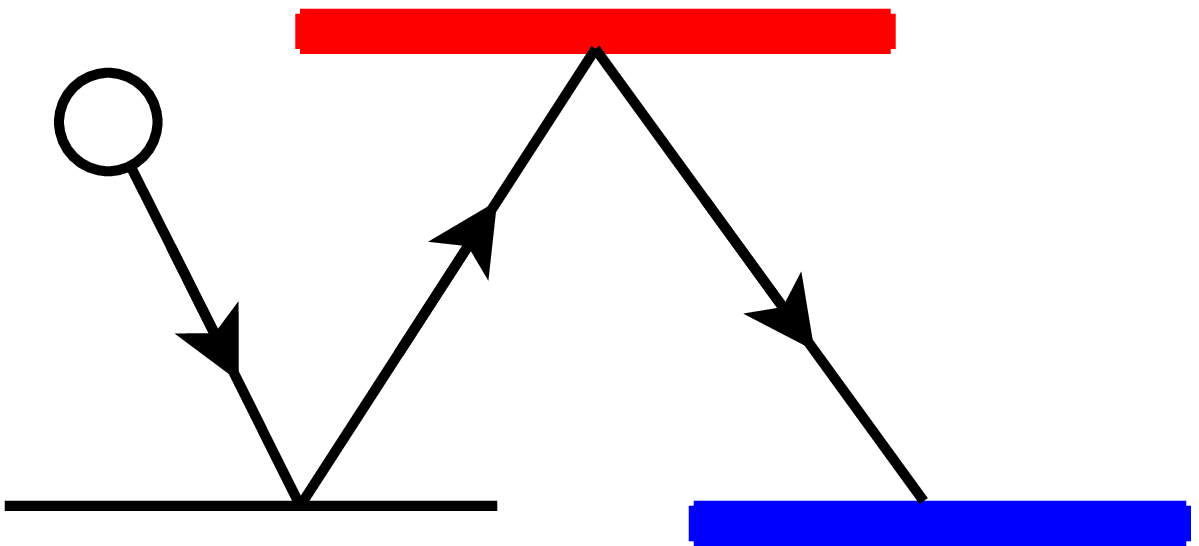} & $-l^j_at_i^ar_j$  \\
    &\includegraphics[scale=0.1]{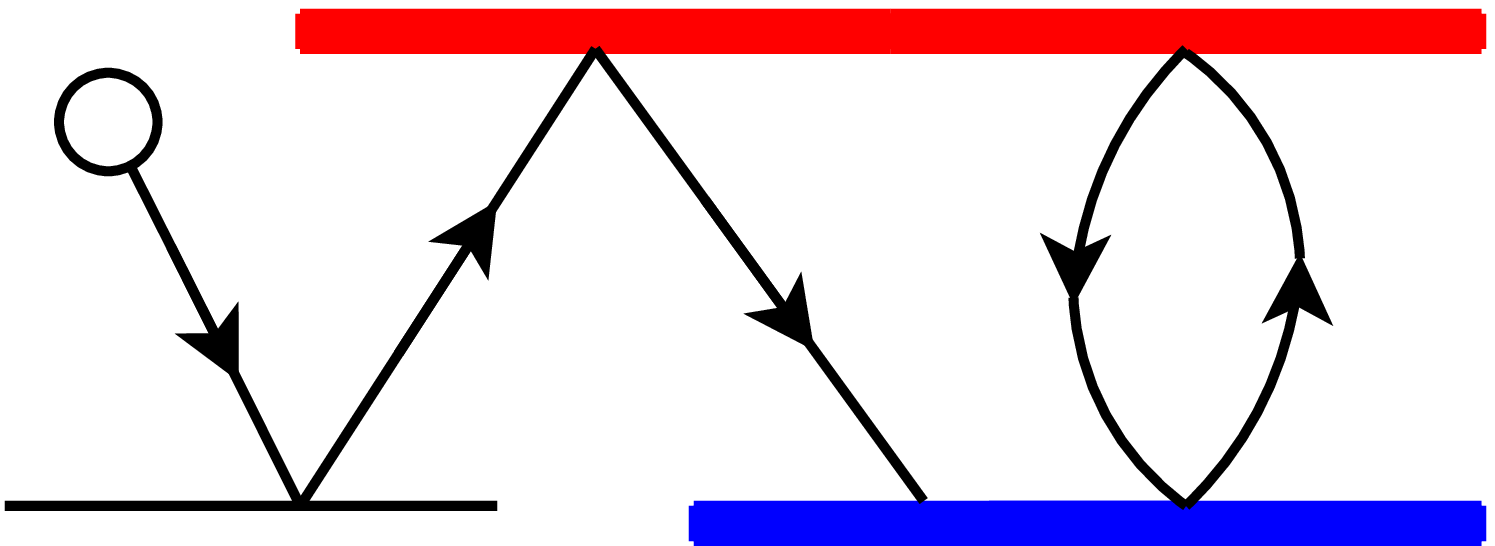} & $-\frac12 l_{ab}^{jk}t_i^ar_{jk}^b$ \\
    &\includegraphics[scale=0.1]{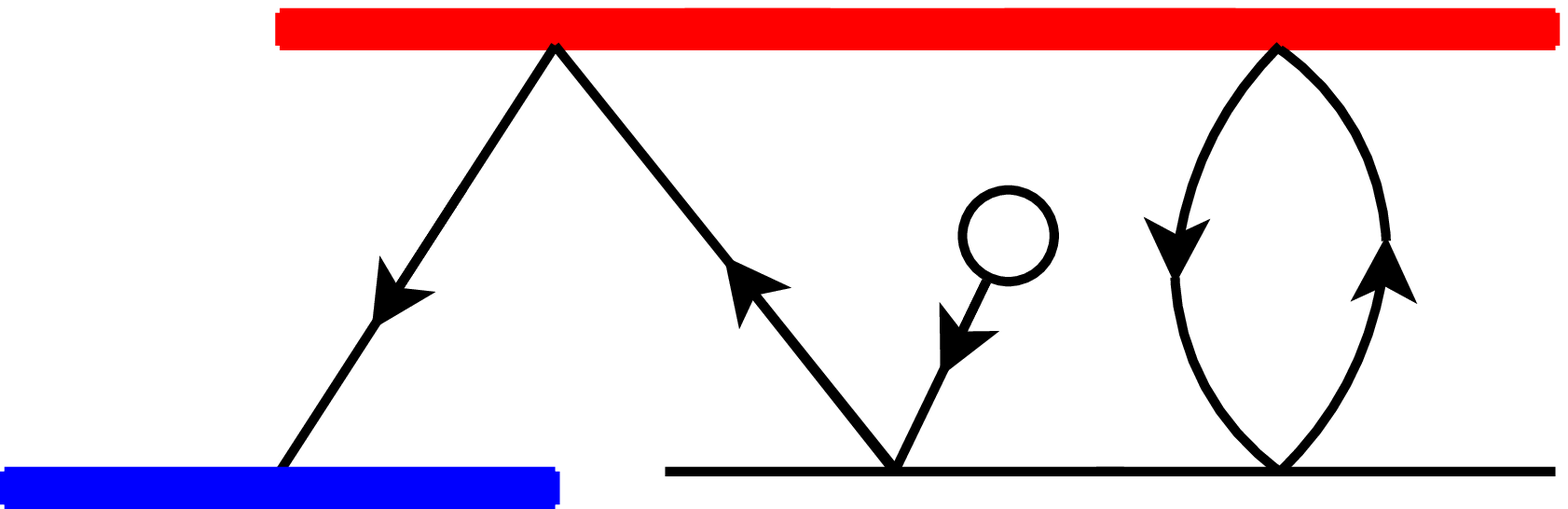} & $-\frac12 l_{ab}^{jk}t_{ik}^{ab}r_j$
  \end{tabular}
  \caption{
    Diagram representation of the overlap expressions; repeated indices implies
    a summation.  If the closed-shell system is in the ground state, diagrams
    involving either  $r_i^a$ or $r_{ij}^{ab}$ vanish.  The individual
    components of these diagrams are explained in Tables \ref{figDiagramsTLR}
    and \ref{figDiagramsOperators}.
  } \label{figDiagramsOverlaps}
\end{table}

\section{Results}
\label{secRes}
In this section we present our results for the calculation of the
spectroscopic factor using \emph{ab initio} coupled-cluster theory.
We study the spectroscopic factor of nucleon removal from $^{16}$O by
calculating the one-body overlap functions of $^{16}$O with the odd mass
neighbors $^{15}$O and $^{15}$N using the PR-EOM-CCSD approach to the ground
and excited states of the $A-1$ nuclei.  The CCSD approximation is used to
calculate the ground state of $^{16}$O.

\begin{figure}[h]
\includegraphics[width=0.45\textwidth,clip=]{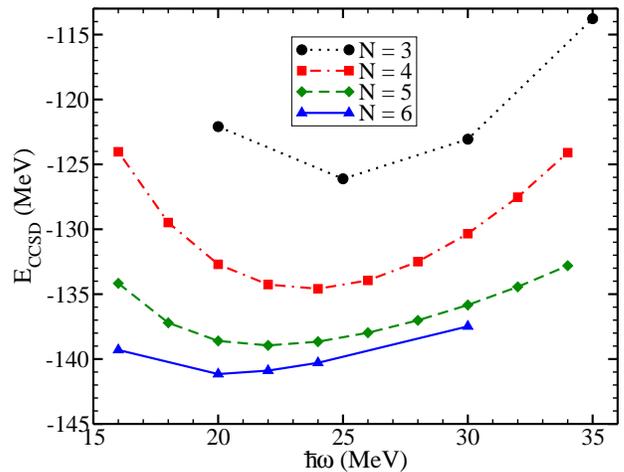}
\caption{(Color online) Convergence of the ground-state energy (within CCSD) of
$^{16}$O using a low-momentum potential with cut-off $\lambda=2.0$fm$^{-1}$
for increasing model space size $N=2n+l$ and as a function of the oscillator spacing $\hbar\omega$.}
\label{figEnergyVShwO16}
\end{figure}

\begin{figure}[h]
\includegraphics[width=0.45\textwidth,clip=]{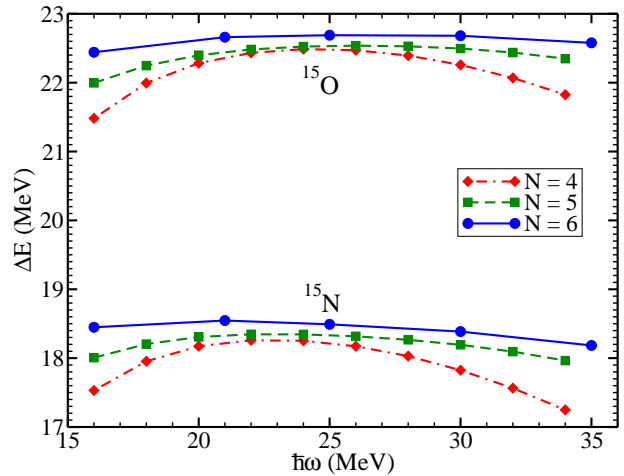}
\caption{(Color online) Convergence of the ground-state energies (within PR-EOMCCSD) of
$^{15}$O and $^{15}$N relative to the ground-state energy of $^{16}$O
using a low-momentum potential with cut-off $\lambda=2.0$fm$^{-1}$
for increasing model space size and as a function of the oscillator spacing $\hbar\omega$.}
\label{figEnergyVShwO15N15}
\end{figure}

Our model space is spanned by oscillator states. We label the model
space by the largest principal quantum number $N$ that is included in
the single-particle (s.p.) basis, so that the maximum s.p. energy is
$E_N = (N+\frac32)\hbar\omega$, and the number of major oscillator
shells is $N+1$.  In Fig.~\ref{figEnergyVShwO16} we show the
convergence of the ground-state of $^{16}$O with increasing size of
the model space for a wide range of oscillator frequencies
$\hbar\omega$, using $V_{\mathrm{low-}k}$ with momentum cutoff
$\lambda=2.0$fm$^{-1}$.  In Fig.~\ref{figEnergyVShwO15N15} we show the
convergence of the ground-state energies of $^{15}$O and $^{15}$N
relative to the ground-state energy of $^{16}$O.

Our results shown in Figs.~\ref{figEnergyVShwO16} and
\ref{figEnergyVShwO15N15} show a weak dependence on $\hbar\omega$ in
the largest model space.  We estimate that our results for the ground
state energies are converged within a few MeV in our largest model.
We note that the CCSD ground-state for $^{16}$O is overbound by $\sim
15$MeV as compared to experiment.  However, the energy difference
between the ground-states of $^{15}$O and $^{15}$N is about $\sim
4$~MeV, which is very close to the experimental value of $3.5$~MeV. It
thus seems that energy differences are better reproduced than absolute
energies.

The $\hbar\omega$ dependence provides some information about how the
finite size of the model space affects the solutions.  For high values
of $\hbar\omega$, the model space includes high-momentum states beyond
the momentum cutoff $\lambda$ of the interaction, but is not
sufficiently extended in position space to accommodate a nucleus.  For
small values of $\hbar\omega$, the model space is sufficiently wide in
position space for the extenson of the nucleus but does not contain
sufficient high-momentum modes to resolve the cutoff $\lambda$ of the
interaction. Close to the minimum, in the largest model spaces
considered, a good compromise is realized.

We also studied the energy levels using $V_{\mathrm{low-}k}$ for
various momentum cut-offs in the range $\lambda=1.6$--$2.2$~fm$^{-1}$.
The calculated ground-state energies for $^{16}$O, $^{15}$O, and
$^{15}$N are sensitive to the cut-off, implying that induced
three-body forces and short-ranged forces of higher rank would
contribute significantly to the calculated energies.

\begin{figure}[h]
\includegraphics[width=0.45\textwidth,clip=]{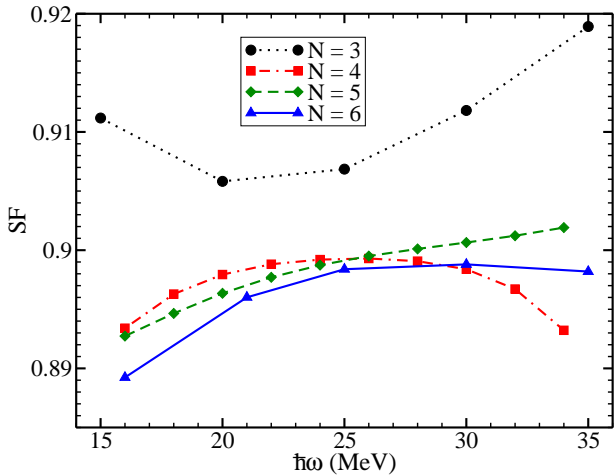}
\caption{(Color online) Spectroscopic factor SF$(1/2^-)$ for proton removal from
  \nuc{16}{O} as a function of the oscillator spacing $\hbar\omega$
  for different model spaces consisting of $(N+1)$ oscillator shells
  and a low-momentum interaction with cutoff $\lambda=2.0$~fm$^{-1}$.}
\label{figSpectroscopicFactorsProtonN}
\end{figure}

Let us turn to the spectroscopic factor for nucleon removal from
$^{16}$O.  Figure~\ref{figSpectroscopicFactorsProtonN} shows the
spectroscopic factor~(\ref{equSFdefinitionCC})
\begin{equation}
\label{key}
{\rm SF(1/2^-)}\equiv S^{16}_{15}(l=1,j=1/2)
\end{equation}
for the removal of a proton with quantum numbers $J^\pi=1/2^-$ from
$^{16}$O using a low-momentum interaction $V_{\mathrm{low-}k}$ with a
cut-off $\lambda=2.0$fm$^{-1}$.  Evidently, the spectroscopic factor
is well converged and depends very weakly on the size of the model
space and the oscillator frequency $\hbar\omega$.  It varies less than
1\% over a wide range of oscillator frequencies. The spectroscopic
factor SF$(1/2^-)$ for neutron removal from $^{16}$O is almost
identical to the SF$(1/2^-)$ for proton removal. Recall that isospin
is approximately conserved in light nuclei.

\begin{figure}[h]
  \includegraphics[width=0.45\textwidth,clip=]{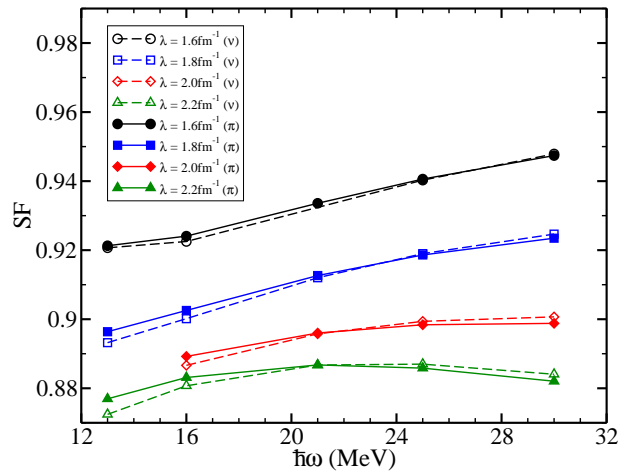}
  \caption{(Color online) Spectroscopic factor SF$(1/2^-)$ for neutron and proton
    removal as a function of the oscillator spacing $\hbar\omega$ for
    nucleon-nucleon interactions with different cutoffs in a model
    space with $N=6$.}
  \label{figSpectroscopicFactorsNeutronProton}
\end{figure}

The dependence on momentum cut-off $\lambda$ is displayed in
Fig.~\ref{figSpectroscopicFactorsNeutronProton}.
Note that the spectroscopic factor increases with decreasing cutoff.
This is expected, since by lowering the cutoff the system becomes less
correlated and the product state $|\phi_0\rangle$ becomes an increasingly good
approximation, and the single-particle picture
becomes more and more valid. Note also that isospin is approximately a
good quantum number, as the spectroscopic factors for proton and
neutron removal are almost identical.

\begin{figure}[h]
\includegraphics[width=0.45\textwidth,clip=]{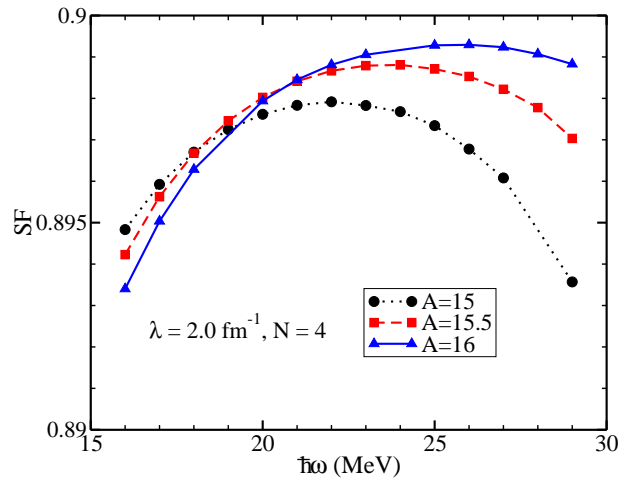}
\caption{(Color online) Spectroscopic factor SF$(1/2^-)$ for proton
  removal from $^{16}$O as a function of the oscillator spacing
  $\hbar\omega$ computed for different values of the mass number $A$
  employed in the intrinsic Hamiltonian~(\ref{equHamiltonOperatorA}).
  The model space consists of $N+1=5$ oscillator shells, and the
  momentum cutoff of the nucleon-nucleon interaction is
  $\lambda=2.0$~fm$^{-1}$.}
\label{figAdependence}
\end{figure}

Let us also study the center-of-mass problem. The intrinsic
Hamiltonian~(\ref{equHamiltonOperatorA}) depends on the mass number
$A$ of the nucleus, and the calculation of the spectroscopic factor
requires us to employ identical Hamiltonians for the nuclei with mass
numbers $A$ and $A-1$. This constitutes dilemma, since no choice of
actual value for the parameter $A$ can satisfy the parent and daughter
nuclei simultaneously.  It is thus necessary to investigate how
strongly the spectroscopic factor depends on this value.
Figure~\ref{figAdependence} shows the spectroscopic factor (in a model
space $N=4$ for a momentum cutoff $\lambda=2.0$~fm$^{-1}$ for
different values of the mass number $A$ of the intrinsic Hamiltonian.
The dependence on $A$ is very weak, and it is similar in size to the
dependence on the parameters of the model space.

For an \emph{intrinsic} Hamiltonian, the coupled-cluster wave function
of a closed-shell nucleus factorizes into an intrinsic part and
Gaussian for the center of mass of coordinate~\cite{HPD09}.  Following
the procedure of Ref.~\cite{HPD09}, we confirmed that this
factorization is present for the ground states of $^{15}$O and
$^{15}$N in the largest model space we considered. We found that this
factorization even takes place if the value $A=16$ for the mass number
is employed in the intrinsic Hamiltonian~(\ref{equHamiltonOperatorA})
for the computation of the nuclei $^{15}$O and $^{15}$N.  These
results suggest that our approach to calculate spectroscopic factors
within the coupled-cluster method is practically free of any
center-of-mass contamination.

\begin{figure}[h]
\includegraphics[width=0.45\textwidth,clip=]{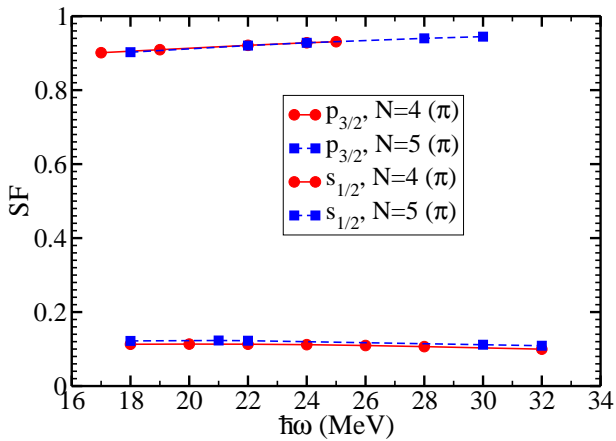}
\caption{(Color online) Spectroscopic factors SF$(3/2-)$ and SF$(1/2^+)$ for
	proton removal from $^{16}$O as a function of the oscillator spacing
	$\hbar\omega$.  The employed model spaces have $N+1$ oscillator shells, and
	the momentum cutoff of the nucleon-nucleon interaction is
	$\lambda=2.0$~fm$^{-1}$.}
\label{figSFp32}
\end{figure}

So far, we focused on the spectroscopic factors for removal of a
$J^\pi=1/2^-$ proton and neutron from $^{16}$O. We finally also
compute the spectroscopic factor for removal of a $J^\pi=3/2^-$ proton
and a (deeply bound) $J^\pi=1/2^+$ proton from $^{16}$O. The result is
shown in Fig.~\ref{figSFp32} for different model spaces. As before,
the results are well converged with respect to the size of the model
space, and display only a mild dependence on the oscillator frequency.
We find that the spectroscopic factor SF$(3/2^-)$ is similar in size
to SF$(1/2^-)$. This is an interesting result. Barbieri and
Dickhoff~\cite{Barbieri2009spe} also found in their computation of
spectroscopic factors that ${\rm SF}(1/2^-)\approx {\rm SF}(3/2^-)$
for nucleon removal from $^{16}$O. As expected, the spectroscopic
factor of the $J^\pi=1/2^+$ state is very small. The removal of a
deeply bound $J^\pi=1/2^+$ proton from $^{16}$O yields a highly
excited state of $^{15}N$ that is a rather complex superposition of
many $n$-particle--$(n+1$)-hole states and thus has little overlap
with a one-hole state.

\section{Conclusion and outlook} \label{secCnO}

We have extended the coupled-cluster method for the computation of
spectroscopic factors.  To this purpose, we derived diagrammatic and
algebraic expressions of the one-body overlap functions based on
equation-of-motion methods for the ground and excited states of the
closed-shell nucleus with mass number $A$ and the neighboring nuclei
with mass number $A-1$. We implemented the equations in an uncoupled
$m$-scheme, and presented proof-of-principle calculations of the
spectroscopic factor for proton and neutron removal from $^{16}$O.
The calculated spectroscopic factors are well converged in model
spaces consisting of six oscillator shells for low-momentum
nucleon-nucleon interactions. Within the coupled-cluster approach, the
same intrinsic Hamiltonian has to be employed in the nuclei with mass
numbers $A$ and $A-1$. We found that the spectroscopic factor is
insensitive to the actual value of the mass number that is employed in
the intrinsic Hamiltonian.

We plan to implement the computation of the spectroscopic factor also
in a spherical formulation of nuclear coupled-cluster theory. This
will allow us to employ much larger model spaces, and we plan to apply
the techniques to physically interesting nuclei, such as $^{22,24}$O,
\nuc{48,52}{Ca}, and \nuc{56,78}{Ni}.

%

\begin{acknowledgments}

	We acknowledge discussions with C.~Barbieri, E.~Bergli, R.~J.~Furnstahl and
	M.~Hjorth-Jensen. \O.~J. thanks the University of Oslo and Oak Ridge National
	Laboratory (ORNL) for hospitality.  This research was partly funded by the
	Norwegian Research Council, project NFR 171247/V30, and by the U.S.
	Department of Energy under grant Nos. DE-FG02-96ER40963 (University of
	Tennessee) and DE-FC02-07ER41457 (SciDAC UNEDF). This research used resources
	of the National Center for Computational Sciences at ORNL.

\end{acknowledgments}

\bibliography{sf_ccm_paper}

\end{document}